\documentclass[preprint,showpacs,preprintnumbers,nofootinbib]{revtex4}
\usepackage{graphicx}
\usepackage{dcolumn}
\usepackage{bm}
\usepackage{amsfonts}

\newcommand{\bea}{\begin{eqnarray}}
\newcommand{\eea}{\end{eqnarray}}
\newcommand{\be}{\begin{equation}}
\newcommand{\ee}{\end{equation}}
\newcommand{\rsdr}{\sqrt{\alpha^{\prime}}}
\newcommand{\sdr}{\alpha^{\prime}}

\begin{document}
\preprint{Brown-HET-1401}
\title{ Moduli Stabilization with the String Higgs Effect}
\author{Scott Watson}
\email{watson@het.brown.edu} \affiliation{Physics Department,
Brown University, Providence RI 02912 USA.}
\date{\today}

\begin{abstract}
We review the notion of the Higgs effect in the context of string theory. We find that by including
this effect in time dependent backgrounds, one is led to
a natural mechanism for stabilizing moduli at points of enhanced gauge symmetry.  We consider this mechanism
for the case of the radion (size of the extra dimensions)
and find that as decompactification of the large spatial dimensions
takes place the radion will remain stabilized at the
self dual radius.  We discuss how this mechanism can be
incorporated into models of string cosmology and brane inflation to resolve some
outstanding problems.  We also address some issues regarding which
string states should be included when constructing low energy actions in string cosmology.
\end{abstract}
\maketitle

\section{Introduction}
Superstring theory predicts the existence of a number of massless scalar fields, known as moduli
fields. These fields are troublesome for both
string theory and string cosmology.
From the string theory perspective they represent flat directions leading to a degeneracy
in the vacuum state for the string, i.e. string theory fails to determine
its own vacuum configuration.  From the perspective of cosmology
these massless scalars are problematic because they would have dramatic
effects on the cosmological evolution, which seem to be in
contradiction with observation.  One resolution
to the so-called cosmological moduli problem is to address these
issues through considering models of string cosmology.
In this way, one may argue that the {\em real} vacuum state is
chosen among the many possibilities by considering the
cosmological evolution.

One attempt at such a construction occurs in
models of string gas cosmology, also known as brane gas
cosmology (BGC)\cite{BGC}.  In these models, in addition to the massless modes
of the string, one includes the massive modes in the form
of a gas of string winding and momentum modes through their stress energy tensor.
One can then show
that the radion, the modulus giving the overall size of the extra
dimensions, is fixed near the string scale by the cosmological
evolution \cite{me1,also}.  Although these results are promising,
it is somewhat inconsistent to include the massive
modes into a low energy effective theory.  Moreover, it was shown
in \cite{eff} that at late times, when the dynamics are described by the $4D$
effective theory, stabilization will only remain for the special case
of one extra dimension.  Both of these problems present serious
challenges for models of BGC.

Another method for fixing moduli has been considered in flux
compactifications of Type IIB string theory \cite{kachru}.  In
these models it is possible to fix nearly all of the moduli by
introducing fluxes wrapping the cycles of the Calabi-Yau.
Moreover, it has been possible to construct slow roll inflation
models in the warped geometry background where the role of the
inflaton is played by the separation of the $D3$ brane in the bulk
and a stack of $\bar{D}3$ branes at an IR fixed point \cite{liam}.
One drawback of these models is the method for stabilizing the
size of the extra dimensions seems to be incompatible with
slow-roll inflation.  It was suggested in \cite{liam} that this
problem might be alleviated by considering additional methods for
stabilizing the radion (size of the extra dimensions).  The method
we present here, motivates low energy corrections that should be
added to the Kahler potential and results in an additional method
for stabilizing the radion.

In this paper we consider the realization of the string Higgs
effect in cosmology.  We take as our starting point the massless
modes of the string as described by the low energy effective
action.  We then find that there exists a critical radius at which
there are additional massless modes.  As one moves away from this
point of enhanced gauge symmetry (ESP) one finds that these
particles gain a mass in a manner analogous to that of electroweak symmetry
breaking.  By including the backreaction of these additional modes on the
evolution, along with the damping of the cosmological expansion,
we find a method for stabilizing the radion.
This is similar to work recently discussed in \cite{stanford},
where in that paper the example was that of a two brane system and
the ESP was related to the inter-brane separation.

In Section II, we discuss the necessary string theory background for
constructing the cosmological model of interest.  We review string compactifications from both
the space-time and string worldsheet conformal field theory (CFT)
perspectives.  We focus on the low energy spectrum and hope to
convince the reader that one needs to be cautious when including
string matter into low energy effective actions.  In particular,
we demonstrate that for a generic radius of compactification, no
winding or momentum states should be present in the spectrum.
We then consider the spectrum near an ESP,
where the gauge theory of the string is enriched by additional massless
states.  This is an example of the Higgs effect in string theory \cite{bagger} and when
combined with a time dependent or cosmological background will result in interesting consequences.
In particular, we demonstrate in Section III
that by considering the additional massless states created at the ESP we can arrive
at a natural mechanism for stabilizing the moduli of the theory.
In this case the modulus of interest will be the radion, i.e.
 the radius of the extra dimensions, and by considering the production of
massless string modes we find a natural way to dynamically determine the
overall size of the extra dimensions. We will conclude with some brief remarks on how this work
could be incorporated into current models of string cosmology.

\section{String Compactifications}
In this section, we review the low energy effective action
for the string, focusing on the bosonic or Neveu-Schwarz (NS) degrees of freedom.
We want to focus in particular on which additional modes could in
principle be added to the low energy action.  We consider the additional
modes that arise from compactification, or perhaps more
appropriately from
decompactification of the large $3+1$ space-time.  We find
that many of these modes become massless near so-called enhanced symmetry points (ESP).
We will see that this results in a natural example of the Higgs effect in string theory \cite{bagger}
and including these modes
back into the effective action results in non-trivial
effects for the evolution of the radion.

We take as our starting point the low energy effective action for the
NS-NS sector of the string in $D=4+d$ space-time dimensions (for a review see \cite{supercosmo}),
\be \label{action1}
S_D=\frac{1}{2 \kappa_0^2}\int d^{4+d}x \sqrt{-G} \; e^{-2 \Phi} \Bigl( R_D+4 (\nabla \Phi)^2 -\frac{1}{12} H^2
+ {\cal O}(\sdr) \Bigr) + {\cal O}(g_s) ,
\ee
where $H_{MNP}=\nabla_{[M}B_{NP]}$ is the three form flux, $\Phi$ is the dilaton, and $G_{MN}$ is the $D$ dimensional
metric with $M,N=0 \ldots D-1$.  The constant $\kappa_0^2$ is at this stage an arbitrary constant that can be redefined
by a shift of the dilaton.  This action represents a double
perturbative expansion in both the string tension $\alpha^{\prime} \sim l_s^2$ and
the string coupling $g_s^2=e^{2\Phi}$.  We should only
trust this action for a large radius of curvature ($l_s^2 \ll R_c$) and
weak coupling ($g_s \ll 1$).  In particular, if we wish
to add string matter to the action we should respect these low energy
and large radius approximations.

From the perspective of the worldsheet
(CFT), the action (\ref{action1}) ensures that the worldsheet couplings $G_{MN}$, $B_{MN}$, and $\Phi$ are scale invariant
by the vanishing of their renormalization group $\beta$ functions
\cite{polchinski}.  These couplings appear in the worldsheet
action which is described by the nonlinear sigma model
\be \label{polyakov}
S_0=\frac{1}{\pi \sdr} \int d^2z \Bigl[ G_{MN}(X)+B_{MN}(X) \Bigr]\partial X^M
\bar{\partial} X^N +\sdr {\cal R}^{(2)} \Phi(X),
\ee
where ${\cal R}^{(2)}$ is the worldsheet Ricci scalar, $X^{M}$ are the
worldsheet fields, and ${\partial}$
($\bar{\partial}$) is the left (right) derivative on the $2d$ worldsheet.
One finds that by
demanding the $\beta$ functions vanish for these couplings, we regain their
usual equations of motion, where we intepret the couplings as the low energy supergravity fields in the
target space.
The CFT approach can be used to calculate S-matrix elements by expanding the
couplings about their classical values,
\bea \label{fluct}
G_{MN}&=& G^{(0)}_{MN} + \tilde{h}_{MN}, \nonumber \\
B_{MN}&=& B^{(0)}_{MN} + \tilde{b}_{MN}, \nonumber \\
\Phi&=&   \Phi^{(0)} + \tilde{\Phi},
\eea
where the perturbations $\tilde{h}_{MN}$, $\tilde{b}_{MN}$, and $\tilde{\Phi}$
represent the specific string states of the graviton, antisymmetric
tensor, and the dilaton, respectively.  One can then include other
string states by inserting additional
vertex operators ${\cal O}(z,
\bar{z})$ with (left, right) conformal dimension (1,1) which
represent marginal deformations of the CFT,
\be
S_0 \longrightarrow S^{\prime}=S_0+\lambda \int d^2z {\cal O}(z,
\bar{z}).
\ee

These marginal deformations
take one CFT $S_0$ into another $S^{\prime}$,
providing a connected set of CFTs that all represent the same physical theory \cite{targetspace}.
This neighborhood around the CFT fixed point makes up the moduli space
and the couplings of these marginal operators are the moduli.
The motion of the moduli are then given by their RG
flow, where $\lambda$ represents the infinitesimal motion.
One problem of string theory is to determine a vacuum expectation value
(VEV)
for each field (or worldsheet coupling), thus fixing these moduli
to some unique value.
If we attempt to do this by including additional matter,
we have argued that this should respect the low energy approximation.
In the CFT, this means
that the additional fields must
correspond
to {\em exactly} marginal deformations.
So we see that by demanding conformal invariance and working in an
effective theory, there are very tight constraints on the
possibilities of adding additional matter.  In particular, if we
are to consider additional string states they should be nearly
massless.  We will see that such states arise after
compactification and at special points in the moduli space.

Motivated by the Brandenberger-Vafa scenario \cite{bv},
we now consider the decompactification of four dimensions
resulting in ${\cal M}^4 \times {\cal K}^d$ where ${\cal K}$ remains
compactified and ${\cal M}^4$ corresponds to an Friedmann-Robertson-Walker (FRW) or time dependent cosmological
background
\footnote{More precisely, in \cite{bv} the geometry was taken as $\mathbb R \times T^3 \times T^6$
where $T^n$ is the $n$ torus.  At late times, when the $T^3$ has grown large the geometry is
approximately ${\cal M}^4 \times {\cal K}^d$ where ${\cal M}^4$ is Minkowski space-time.}.
We assume that the four dimensions have expanded
enough so that the curvature scale of ${\cal M}^4$ is much larger than
the scale of the compact dimensions (i.e. $R^{(4)}_c >>
R_c^{(d)}$).  In this limit, the low energy
degrees of freedom are given by the dimensionally reduced action,
\bea \label{action2}
S=\frac{1}{2 \kappa_4^2}\int d^{4}x \sqrt{-g} \; e^{-2\varphi} \Biggl[ R + 4(\nabla \varphi)^2 +\frac{1}{4}\nabla_{\mu} h^{ab}
\nabla^{\mu} h_{ab}-\frac{1}{4}{\cal F}_{\mu \nu a} {\cal F}^{\mu \nu a}
\nonumber \\
-\frac{1}{12}H_{\mu \nu \lambda}H^{\mu \nu \lambda}-\frac{1}{4}H_{\mu \nu a} H^{\mu \nu a}
-\frac{1}{4}H_{\mu a b} H^{\mu a b} -\frac{1}{12}H_{a b c}H^{a b c} \Biggr],
\eea
where $M, N =0 \ldots (D-1)$ refer to the full space-time,
$\mu, \nu=0 \ldots 3$ denote the ${\cal M}^4$ directions, and $a,b,c=4 \ldots (D-1)$
denote the compact dimensions of ${\cal K}$.
The $D$ dimensional metric is given by,
\be
\label{metric}
{G}_{MN}  =
     \left( \begin{array}{cc}  g_{\mu\nu}+ h_{ab}{A_{\mu}}^{a}
A_{\nu}^{b} & \;\; A_{\mu b} \\ A_{\nu a} & \;\; h_{ab}
\end{array} \right),
\ee
and we have introduced the four dimensional dilaton $\varphi \equiv
\Phi-\frac{1}{2} \ln \det h_{ab}$ and $2\kappa_4^2 = 16 \pi G =2 \pi \alpha^{\prime} g_s^{2}$.
The compactification of $B_{MN}$ follows analogous to
(\ref{metric}) with $B^D_{\mu \nu}=B_{\mu \nu}+h_{a b}B_{\mu}^{a}B_{\nu}^{b}$.
In general, we will be interested in internal metrics of the form $h_{ab}=\sdr e^{2\sigma}
\delta_{ab}$.
\subsection{String Spectrum on $M^4 \times S^1$}
We now consider the case of one extra
dimension with metric $G_{55}=\sdr e^{2\sigma}$, where $\sigma$ is the radion and we approximate
the large dimensions as Minkowski space-time $M^4$.
We make this assumption for
simplicity and note that our main results can be extended to more general
cases, such as the $d=6$
compactification of the Heterotic string.
The action (\ref{action2}) simplifies to
\be \label{action3}
S=\frac{1}{2 \kappa_4^2}\int d^{4}x \sqrt{-g} \; e^{-2\varphi} \Bigl( R + 4(\partial \varphi)^2
- (\partial \sigma )^2-\frac{1}{4}e^{2 \sigma} {\cal F}_{\mu \nu} {\cal F}^{\mu
\nu}-\frac{1}{4}e^{-2 \sigma} B_{\mu \nu} B^{\mu
\nu}-\frac{1}{12}H^2 \Bigr),
\ee
where $H_{\mu \nu \lambda}=\partial_{[\mu}B_{\nu \lambda]}$ is the flux in four dimensions and ${\cal F}_{\mu \nu}$
contains the usual Kaluza-Klein (KK) vectors of mass
$m^2=\frac{n^2}{R^2}$ resulting from the compactified metric i.e., $G_{\mu 5}$.
In addition, we also get another $U(1)$ from the
compactification of the higher dimensional antisymmetric two form
$B_{\mu 5}$ resulting in the winding states with $m^2=\frac{\omega^2 R^2}{\alpha^{\prime
2}}$.

We can also see these states arising from the more fundamental worldsheet action
(\ref{polyakov}). Assuming the couplings don't depend on the extra
dimension we find that the reduced action becomes
\bea \label{p4}
S_{4p}&=&\frac{1}{\pi \sdr} \int  d^2z \Bigl[ G_{\mu \nu}(X)+B_{\mu \nu}(X) \Bigr]\partial
X^{\mu} \bar{\partial} X^{\nu}
+\Bigl[ G_{\mu 5}(X)+B_{\mu 5}(X) \Bigr] \bar{\partial} X^{\mu}
{\partial} X^{5}
\nonumber \\
&&+\Bigl[ G_{\mu 5}(X)-B_{\mu 5}(X) \Bigr] {\partial} X^{\mu}
\bar{\partial} X^{5}
+ G_{5 5}(X) \partial X^{5}
\bar{\partial} X^{5}
+\sdr {\cal R}^{(2)} \Phi(X),
\eea
where $R \equiv G^{1/2}_{55}=\rsdr e^{\sigma}$ is the radius of the extra
dimension.
To manifest the chiral symmetry (right and left currents move independently on the string), we can define
\bea
A_{\mu}\equiv A_{\mu}^{Left}=\frac{1}{2} \Bigl( G_{\mu 5}+B_{\mu 5} \Bigr), \nonumber \\
\bar{A}_{\mu}\equiv A_{\mu}^{Right}=\frac{1}{2} \Bigl( G_{\mu 5}-B_{\mu 5} \Bigr).
\eea

The mass of a string state
is found by enforcing
conformal symmetry mode by mode and results in the Virasoro constraints \cite{polchinski},
\bea \label{mass}
M^2=\frac{n^2}{R^2}+\frac{\omega^2 R^2}{\alpha^{\prime
2}}+\frac{2}{\alpha^{\prime}}(N+\bar{N}-2), \nonumber \\
n\omega+N-\bar{N}=0,
\eea
where the integers $n$ and $\omega$ label the momentum and winding
charge and $N$ ($\bar{N}$) correspond to the number of left (right) oscillators that are excited.
We are interested in the massless states and we can see from (\ref{mass})
that for generic radii this
means we are restricted to the states
\be
n=\omega=0 \;\;\;\;\;\; N=\bar{N}=1.
\ee

The effective lagrangian representing the additional degrees of
freedom from the string compactification takes form
\be \label{lmatter}
{\cal L}_m=(\partial \sigma)^2-\frac{1}{4g^2}(F_{\mu \nu}F^{\mu \nu})
-\frac{1}{4{g}^2}(\bar{F}_{\mu \nu}\bar{F}^{\mu \nu}),
\ee
with
\be \label{fields}
F_{\mu \nu}=\partial_{[ \mu} A_{\nu]} \;\;\;\;\;\;\; \bar{F}_{\mu \nu}=\partial_{[ \mu}
\bar{A}_{\nu]},
\ee
and the scalar is neutral and comes from $N=\bar{N}=1$ being taken in the compact direction.
Thus, we have a $U_L(1)
\times U_R(1)$ chiral gauge theory resulting from the compactification, along
with the neutral scalar $\sigma$ giving the overall scale of the extra dimension.
This doubling of the usual KK spectrum is an example of how
strings lead to an enrichment of the usual symmetries.
In summary, we have found that the low energy physics is described by dilaton gravity coupled to a $U_L(1) \times
U_R(1)$ chiral gauge theory with a neutral scalar field.

As in the uncompactified case, we are interested in the fluctuations about the
background values.  For the new states resulting from
compactification we have the lower dimensional analogs of
(\ref{fluct}) along with the new states
\bea
A_{\mu}&=&A_{\mu}^{(0)}+A_{\mu}^{(3)},\nonumber \\
\bar{A}_{\mu}&=&\bar{A}_{\mu}^{(0)}+\bar{A}_{\mu}^{(3)},\nonumber \\
G_{55}&=&\sdr(1+\phi^{(33)}),
\eea
where the first term on the right is the background value.
Our notation for the left (right) going vector perturbation
$A^{(3)}_{\mu}$ ($\bar{A}_{\mu}^{(3)}$) and scalar perturbation $\phi^{(33)}$ will
become clear shortly.  The vertex operators corresponding to the
insertion of these states are given by
\bea \label{mm}
V^{(3)}=A^{(3)}_{\mu} \partial X^5 \bar{\partial} X^{\mu}, \nonumber \\
\bar{V}^{(3)}=\bar{A}^{(3)}_{\mu} \partial X^{\mu} \bar{\partial}
X^5, \nonumber \\
V^{(33)}=\phi^{(33)} \partial X^5 \bar{\partial} X^5.
\eea
Note that these operator insertions correspond to the
same states that appeared in the
space-time matter lagrangian (\ref{lmatter}).
We can identify $\phi^{(33)}$ with a perturbation of the compact radius away from the
self dual radius and then $\phi^{(33)}=0$ corresponds to $R=\rsdr$.
For $\sigma \ll 1$ we find $\sigma \sim \phi^{(33)} \sim {\delta R}$
gives the departure from the self dual radius.

Let us consider the simple string background $G_{\mu \nu}=\eta_{\mu \nu}$,
$B_{\mu \nu}=0$, and $\Phi=\Phi_0$.
As we have discussed, the low energy approximation requires the fluctuations in
(\ref{mm}) to correspond to
exactly marginal operators and be primary fields of dimension $(1,1)$.
This requirement implies that they must obey the following equations \cite{bagger},
\be
\Box \phi^{(33)}=\Box A^{(3)}_{\mu}=\Box \bar{A}^{(3)}_{\mu}=0 \;\;\;\;\;
\partial^{\mu} A^{(3)}_{\mu}=\partial^{\mu} \bar{A}_{\mu}=0.
\ee
Once again we see that by demanding conformal invariance we obtain
the expected equations of motion for
the fields that would have followed from
({\ref{lmatter}) with the covariant choice of the Lorentz gauge.
We also see from these equations, it is clear that there is nothing to determine
the VEV of $\phi^{(33)}$.  This is a simple example of the
inability of string theory to predict a unique vacuum of the
theory.  That is, each value of $\langle \phi^{(33)} \rangle$ corresponds to a
different choice of vacuum for the {\em same} string theory. From
the perspective of cosmology this implies that the radion
$\phi^{(33)}$ is not stabilized in the low energy theory.
This can be understood from ({\ref{lmatter}), where we saw that with
$n=\omega=0$, the radion
$\phi^{(33)}$ is uncharged under the $U_L(1) \times U_R(1)$ and there is nothing
to fix the scale.  One might expect
that including the winding and momentum modes could change this
result, but we have seen that restricting to the low energy spectrum
eliminates states that carry winding and momentum charge. We will now see that this does
not necessarily remain true if we consider regions of the moduli space near special radii,
so-called ESPs.

\subsection{Enhanced Gauge Symmetry}
Upon examining the mass spectrum given by (\ref{mass}), we see
that at the special radius $R=\rsdr$ there is a possibility of
additional massless states.

These correspond to four new vectors,
\bea \label{vectors}
N=0 & \bar{N}=1 & n=\omega=\pm 1 \;\;\;\;\;\;\;\;\;
V^{(\pm)}={A}_{\mu}^{(\pm)} \bar{\partial}X^{\mu} \exp \Bigl( {\pm
i\frac{2}{\rsdr} X^5}\Bigr),\nonumber \\
N=1 & \bar{N}=0 & n=-\omega=\pm 1 \;\;\;\;\;\;
\bar{V}^{(\pm)}=\bar{A}_{\mu}^{(\pm)} {\partial}X^{\mu} \exp \Bigl( {\pm
i\frac{2}{\rsdr} \bar{X}^5}\Bigr),
\eea
four massless scalars, where $N$ or $\bar{N}$ are taken in the compact direction
\bea \label{scalars1}
N=0 & \bar{N}=1 & n=\omega=\pm 1 \;\;\;\;\;\;\;\;\;\;\;\;
V^{(\pm 3)}={\phi^{(\pm 3)}} \bar{\partial}X^{5} \exp \Bigl( {\pm
i\frac{2}{\rsdr} {X}^5}\Bigr),\nonumber \\
N=1 & \bar{N}=0 & n=-\omega=\pm 1 \;\;\;\;\;\;\;\;\;
V^{(3\pm)}={\phi^{(3\pm)}} {\partial}X^{5} \exp \Bigl( {\pm
i\frac{2}{\rsdr} \bar{X}^5}\Bigr),
\eea
and four massless scalars that are purely winding and purely
momentum charged states
\bea \label{scalars2}
N=\bar{N}=\omega=0 & n=\pm 2 \;\;\;\;\;\;\;\;\;
V^{(\pm \pm)}={\phi^{(\pm \pm)}} \exp \Bigl( {\pm
i\frac{2}{\rsdr} {X}^5}\Bigr) \exp \Bigl( {\pm
i\frac{2}{\rsdr} {\bar{X}}^5}\Bigr), \nonumber \\
N=\bar{N}=n=0 & \omega=\pm 2 \;\;\;\;\;\;\;\;\;
V^{(\pm \mp)}={\phi^{(\pm \mp)}} \exp \Bigl( {\pm
i\frac{2}{\rsdr} {X}^5}\Bigr) \exp \Bigl( {\mp
i\frac{2}{\rsdr} {\bar{X}}^5}\Bigr).
\eea
This concludes the massless spectrum at the ESP $R=\rsdr$.  We see that it
is now consistent to include states with nontrivial winding and momentum into
the low energy spectrum.  These additional vectors (\ref{vectors})
fill out the adjoint representation of $SU_L(2) \times SU_R(2)$ enhancing the
previous symmetry from $U_L(1) \times U_R(1)$.
The eight scalars
(\ref{scalars1}) and (\ref{scalars2}) combine with $\phi^{(33)}$
to transform as the ({\bf 3},{\bf 3}) adjoint representation of the gauge group.
The T-duality that is usually associated with winding/momentum states
can now be realized as a gauge rotation by $\pi$ in one of the $SU(2)$'s
which sends $\phi^{(33)}$ to $-\phi^{(33)}$
or $R$ to $\sdr/R$.  Thus, T-duality arises from the action of a $\mathbb{Z}_2$ subgroup of the
full $SU_L(2) \times SU_R(2)$ symmetry.  This
is an example of the fact that duality holds in the nonperturbative
regime, since the gauge theory approach will remain valid there \cite{targetspace}.

The requirement that these fields maintain conformal invariance, being primary operators
of dimension (1,1), is
\be
\Box A^{(i)}=\Box \bar{A}^{(i)}=\Box \phi^{ij}=0 \;\;\;\;
\partial^{\nu} A_{\nu}^{i}=\partial^{\nu} \bar{A}_{\nu}^{i}=0,
\ee
where $i,j=+,-,3$.  These are the expected equations resulting
from (\ref{lmatter}) after lifting the theory to $SU_L(2) \times
SU_R(2)$ and restricting to the self dual radius, i.e. $\langle
\phi^{(33)}\rangle=0$.
The field strength
(\ref{fields}) is now given by the Yang-Mills theory
\be
F^i_{\mu \nu}=\partial_{\mu} A^i_{\nu}-\partial_{\nu} A^i_{\mu}+
g\epsilon^{ijk}A^j_{\mu}A^k_{\nu},
\ee
\be
\bar{F}^i_{\mu \nu}=\partial_{\mu} \bar{A}^i_{\nu}-\partial_{\nu} \bar{A}^i_{\mu}+
g\epsilon^{ijk}\bar{A}^j_{\mu}\bar{A}^k_{\nu},
\ee
and the scalars couple through the ({\bf i},{\bf 0}) and ({\bf 0},{\bf j}) gauge covariant derivatives
\be \label{gcd}
(D_{\mu}\phi)^a=\partial_{\mu}\phi^a+g\epsilon^{abc}A^b_{\mu}\phi^c,
\ee
\be \label{gcd2}
(\bar{D}_{\mu}\phi)^a=\partial_{\mu}\phi^a
+g\epsilon^{abc}\bar{A}^b_{\mu}\phi^c,
\ee
where the coupling $g$ is of ${\cal O}(1)$ for the states we are considering\footnote{For example, for the heterotic
string the four dimensional gauge coupling is given by $g^2=4 \kappa^2 / \sdr$, where $\kappa$ is the gravitational
length and contains the dilaton expectation value.  One can usually choose these values so that $g$ is order one, which
is expected from the Yang-Mills theory.  This implies that the string scale is close to the gravitation scale.  For a
complete discussion see \cite{polchinski}.}.
\subsection{Higgs Mechanism in String Theory}
We have seen that at the ESP (i.e. $R=\rsdr$) the gauge symmetry is
enhanced to $SU_L(2) \times SU_R(2)$.
Let us consider the mass of the states (\ref{vectors}), (\ref{scalars1}), and
(\ref{scalars2}) away from $R=\rsdr$.  We find
\be \label{espmass}
M=\frac{|R^2-\sdr|}{R\sdr}=\frac{2}{\rsdr}| \sinh(\sigma)|,
\ee
where we have used $R=\rsdr e^{\sigma}$.  Thus, we see that
if the radion has a vanishing VEV these states remain massless,
whereas if $\phi^{(33)} \neq 0$ the states become massive.  This is an
example of the Higgs Mechanism in string theory, where the role of
the Higgs is played by the radion $\phi^{(33)}$.  We can see the Higgs
mechanism explicitly from (\ref{gcd}) and (\ref{gcd2}).  Consider
the ({\bf i},{\bf 0}) covariant derivative for the scalar $\phi^{(+ 3)}$,
\bea
D_{\mu}\phi^{(+3)}&=&\partial_{\mu}\phi^{(+3)} +
g\epsilon^{+bc}A^b_{\mu}\phi^c,\nonumber \\
&=&\partial_{\mu}\phi^{(+3)} +
g\Bigl(A^{-}_{\mu}\phi^{(33)}-A^3_{\mu}\phi^{(-3)}\Bigr).
\eea
As usual, we can perform a gauge rotation to fix
the VEV of $\phi$ in the $({\bf 3},{\bf 3})$ direction $\langle \phi^{(33)}\rangle=\delta
R $ and
through the kinetic term this gives a mass to the vector $A^{(-)}_{\mu}$,
\be \label{massgen}
\frac{1}{2} (D_{\mu}\phi^{(+ 3)})^2 \longrightarrow \frac{1}{2}g^2
\langle \phi^{(33)}\rangle^2  (A^-_{\mu})^2=\frac{1}{2}g^2
\delta R^2 (A^-_{\mu})^2.
\ee
Through this and the other covariant derivative terms the four vectors $A^{\pm}$ and
$\bar{A}^{\pm}_{\mu}$ get masses through their
couplings to $\phi^{(33)}$, while the $A_{\mu}^{(3)}$ and $\bar{A}_{\mu}^{(3)}$ remain massless.
Thus, we see that the $SU_L(2) \times SU_R(2)$ is broken to
$U_L(1) \times U_R(1)$.  As we have discussed, it is problematic
in string theory to determine $\delta R^2$, since no potential
exists for the scalars $\phi$.  However, we
argue in what follows that by considering the time dependence of
the background, one is naturally led to a cosmological mechanism
that breaks this degeneracy.

\section{\label{cosmo} An Effective Potential for Radion Stabilization}
We saw in the last section that the Higgs mechanism arises
naturally in
string theory near ESPs.
In the simple $S^1$ compactification we considered, we found that the low energy
degrees of freedom evolved in accordance with dilaton gravity coupled to a chiral gauge theory $U_L(1) \times
U_R(1)$.  This was true at generic radii and we found that there was no scale to
fix the VEV of the radion.  We also found that near the ESP (i.e.
$R=\rsdr$)
extra states (\ref{vectors}), (\ref{scalars1}), and (\ref{scalars2})
became massless and we should include these states and their interactions
in the effective theory \footnote{Similar considerations were addressed from the
supergravity perspective for the
case of the
heterotic string on $K3 \times S^1$ and $K3 \times T^2$
in \cite{mohaupt1} and \cite{mohaupt2}, respectively.}.

This suggests the following model for stabilizing the radion near the ESP.
Initially the radion $\sigma$ begins
at a generic point in the moduli space and the correct theory is that of dilaton gravity coupled to the
$U_L(1) \times U_R(1)$ massless degrees of freedom.  We assume the
dilaton has been fixed\footnote{Since we are interested in late time cosmology, we will assume
that the dilaton has already been stabilized.  We anticipate this
could be done either by fluxes, wrapped branes, or perhaps even by a variation of the mechanism
we will describe.} and we
consider the late time cosmology in an homogeneous and isotropic universe with metric,
\be
ds_4^2=-dt^2+e^{2\lambda(t)}d\vec{x}^2,
\ee
where $\lambda(t)=\ln a(t)$ is the scale factor.
The effective action for generic $\sigma$ is given by
\be \label{theaction}
S_{eff}=\int d^4x \sqrt{g}\; \Bigl[ R- \frac{1}{2}(\partial
\sigma)^2
- V_{eff} \Bigr],
\ee
where $V_{eff}$ initially represents the contribution from the chiral
$U(1)$s, although near the self dual radius it should incorporate the
effects due to the additional massless states.

Let us consider the background equations of motion first neglecting
the backreaction near the ESP.  The equations following from (\ref{theaction}) are
\bea \label{eqn1}
3\dot{\lambda}^2=\frac{1}{2}\dot{\sigma}^2+\rho_{sub}
\\
2\ddot{\lambda}+3\dot{\lambda}^2=-\frac{1}{2}\dot{\sigma}^2-p_{sub}
 \\ \label{eqno1}
\ddot{\sigma}+3 \dot{\lambda} \dot{\sigma} =\frac{\partial V_{eff}}{\partial \sigma},
\eea
where $\rho_{sub}$ and $p_{sub}$ represent the subdominant contribution from
the $U_L(1) \times
U_R(1)$ contained in $V_{eff}$ at generic radii.
This contribution will be subdominant at early times, since the
kinetic term has an equation of state
$\rho=p$ and thus scales as $\rho=a^{-6}$.  The corresponding scale factor is $a(t) \sim t^{1/3}$
and $\dot{\lambda}=1/3t$.  In this limit we can ignore the potential in (\ref{eqno1})
and $\sigma$ is given for small $t$ as
\be
\sigma(t)=\sigma_0+v_0 t.
\ee
We start the time
evolution at $t=0$ when the field is closest to $R=\rsdr$, thus we see that $\sigma_0$ is a measure of how
close the radion comes to the ESP.  In \cite{me1}, it was shown that
by including the dilaton in the dynamics, along with the winding
and momentum modes of the string, the radion will naturally pass
through $\sigma=0$ and be localized around the ESP\footnote{This result was for the full $10D$ theory in the
string frame.  We note that this result came from trusting the low energy effective action
while including the massive modes of the string. This seems to be an issue that requires further attention.}.
Motivated by this result we assume that
$\sigma_0=0$, which is the most efficient case for particle production, since the
states will be exactly massless there.

We proceed to address particle creation in a way analogous to
the treatment in models of inflationary preheating \cite{preheat}, where
here the role of the inflaton will be played by the radion.  Since we
are really discussing the creation of strings, one might wonder if
we are justified in taking this approach.
It was shown in \cite{production}
that string production can be considered in this way.  There it
was shown that it is enough to consider the production
of strings, mode by mode, in the effective field theory approach.
Using this approach we can think of each string mode as a scalar field
with a time varying mass.

For example, let us consider the effects of producing one of the additional massless
vectors that appears at the ESP.  From the coupling in (\ref{massgen}) we see that
the additional states would lead to a potential
\be \label{thepotential}
V_{eff}(\sigma,A_{\mu})=\frac{1}{2}(\partial_{\mu} A_{\nu})^2-\frac{1}{2} g^2 \sigma^2 A_{\mu}A^{\mu},
\ee
where we have defined $A_{\mu} \equiv A^{(-)}_{\mu}$ and we work in the Lorentz gauge $
\partial^{\mu} A_{\mu}=0$.  Note that we are neglecting the other Yang-Mills
interactions, as these would lead to the same generic dynamics for $\sigma$.
However, it would be interesting to include these
interactions in future work, as they are examples coming directly from string theory
of the type of interactions recently considered in \cite{peebles} as
dark matter candidates.

From (\ref{thepotential}), we can identify $m(t)^2=g^2 \sigma^2$ as a time dependent mass for $A_{\mu}$.
As $\sigma$ approaches the ESP, the $A_{\mu}$'s become massless and easy to create.
Then, as $\sigma$ leaves the ESP
these states will become massive resulting in backreaction and producing an attractive force
pulling $\sigma$ back towards the ESP.

Let us consider the time dependent frequency of a particular Fourier mode $A^{\mu}_k$
\be
\omega_k(t)=\sqrt{\vec{k}^2+g^2 \sigma^2(t)}.
\ee
A particular mode becomes excited when the non-abiabaticity
parameter satisfies ${\dot{\omega}}/{\omega^2} \geq 1$.
When this condition holds for a particular mode, it results in particle production
and an occupation number
\be
n_k=\exp{ \Biggl( -\frac{\pi \vec{k}^2+g^2 \sigma_0^2}{g v_0}} \Biggr).
\ee

Recall that we can take $\sigma_0=0$ and
$g$ in our case is a positive constant of order unity coming from the coupling of the
winding and momentum modes.
The energy density of produced particles is given by
\be
\rho_{A} =\int \frac{d^3k}{(2\pi)^3}n_k \omega_k \approx g | \sigma(t) | N,
\ee
with $N \sim (gv_0)^{3/2}$.  Thus, comparing this to (\ref{eqn1}) we see that the initial kinetic
energy associated with the radion $\frac{1}{2} v_0^2$
is dumped into production of $A_{\mu}$ particles as the radion passes
through the ESP.  Given a large enough $v_0$, the radion will continue its trajectory
and the modes will become massive as we have seen.  This results in an always attractive force of
magnitude $gN$ pointing the radion back towards the ESP.
The effective equation for $\sigma$ including
the backreaction is then given by
\bea \label{eqn2}
\ddot{\sigma}+3 \dot{\lambda} \dot{\sigma} =-g N(t).
\eea
This process will continue with each pass of the radion, until all
of its initial kinetic energy
has been used up and it settles to the
self dual radius.  Therefore, we are led to the conclusion that the additional states
associated with the enhanced symmetry result in a fixed value for
the radion at the self dual radius.

One immediate concern might be whether this method is stable to
perturbations.  Moreover, one could worry that the initial kinetic
energy of the radion is so high that the force associated with the
backreaction is not enough to over come its inertia.  Both of
these problems are overcome by considering the Hubble friction
associated with the second term in (\ref{eqn2}).  One expects this friction
to damp out any perturbations and should actually enhance the stabilization mechanism.
This was discussed in models of string gas cosmology \cite{eff} and a similar conclusion was reached in
\cite{stanford}.  Moreover, it was shown in
\cite{eff} that
once we switch to the effective theory the Hubble friction is enough to keep the
radion evolving slowly compared to the growth of the three large
dimensions.  We conclude that Hubble friction combined with the
ESP backreaction should be more than adequate to stabilize the
radion at the self dual radius.

\section{Conclusions}
We have found that by considering the string Higgs effect it is
possible to generate a potential for the radion resulting from
the backreaction of the additional low energy states that emerge at the
ESP.  These low energy states carry both winding and momentum
charge, justifying their inclusion in low energy effective actions where the
dimensions are taken initially to be near the self dual radius.  This includes models of
Brane Gas Cosmology, which address the issue of stabilization
of extra dimensions from the higher dimensional perspective \cite{me1,also}.
This complements our approach here, where we have demonstrated the stabilization from the lower
dimensional perspective without the help of the dilaton.

When considering the inclusion of these additional modes we have neglected much of the
richness of the nonabelian gauge theory.  That is, we focused on
the coupling of the vectors and radion arising from the covariant
derivatives in (\ref{lmatter}), but we have neglected the
interactions and self interactions of the fields.  Including these additional
interactions would be interesting
for a number of reasons.  They offer a concrete way to include the
interactions of winding and momentum modes into models of Brane Gas Cosmology.
Including additional interactions would also offer an example, coming directly from
string theory, of the dark matter candidates recently proposed in
\cite{peebles}.

Finally, it would be interesting to see if this mechanism can be
combined with models of flux compactifications. As we mentioned in
the introduction, these compactifications can fix all the moduli
in principle, however the fixing of the overall scale of the extra
dimensions seems incompatible with inflation \cite{liam}. Since
this is precisely the radion, in principle, the two methods could
be combined to fix all the moduli.  One problem that prevents this
from immediately being realized is that in these models one usually works
in Type IIB theory and the compactification manifold
is taken to be of the Calabi-Yau type.  These manifolds are chosen because they
offer the
most realistic particle spectrum for Type IIB theories.  In addition, they
possess $SU(3)$ holonomy, which implies that the low energy string spectrum does not admit one cycles.
As a result there is not an ESP corresponding to the self-dual radius and the methods
we have suggested in this paper need not apply.  However, our methods do readily apply to toriodal
compactifications of the Heterotic string and work is
currently underway to explore the additional role of fluxes.  It
will be interesting to see if slow-roll brane inflation can be realized in such a picture.

In closing, we hope the reader has been left with the idea that enhanced
symmetry combined with cosmological evolution presents a viable mechanism for
fixing moduli, but is in need of further study.
\begin{acknowledgments}
I would especially like to thank Steve Gubser for the initial
motivation for undertaking this project. I would also like to
thank Antal Jevicki, Daniel Kabat, David Lowe, Liam McAllister,
and Cumrun Vafa for useful discussions, and Thorsten Battfeld,
Robert Brandenberger and Sera Cremonini for comments on the
manuscript. Financial support was provided by NASA GSRP.
\end{acknowledgments}


\begin{thebibliography}{99}
\bibitem{BGC}
For foundational ideas see: \\
R.~H.~Brandenberger and C.~Vafa,
``Superstrings In The Early Universe,''
Nucl.\ Phys.\ B {\bf 316}, 391 (1989);\\
A.~A.~Tseytlin and C.~Vafa,
``Elements of string cosmology,''
Nucl.\ Phys.\ B {\bf 372}, 443 (1992)
[arXiv:hep-th/9109048];\\
S.~Alexander, R.~H.~Brandenberger and D.~Easson, ``Brane gases in
the early Universe,'' Phys.\ Rev.\ D {\bf 62}, 103509 (2000)
[arXiv:hep-th/0005212];\\
R.~Easther, B.~R.~Greene, M.~G.~Jackson and D.~Kabat,
``Brane gas cosmology in M-theory: Late time behavior,''
Phys.\ Rev.\ D {\bf 67}, 123501 (2003) [arXiv:hep-th/0211124].\\
For later generalizations see \cite{eff} and references within.

\bibitem{me1}
S.~Watson and R.~Brandenberger,
``Stabilization of extra dimensions at tree level,''
JCAP {\bf 0311}, 008 (2003)
[arXiv:hep-th/0307044].

\bibitem{also}
S.~P.~Patil and R.~Brandenberger,
``Radion stabilization by stringy effects in general relativity and dilaton
gravity,''
arXiv:hep-th/0401037;\\
B.~A.~Bassett, M.~Borunda, M.~Serone and S.~Tsujikawa,
``Aspects
of string-gas cosmology at finite temperature,'' Phys.\ Rev.\ D
{\bf 67}, 123506 (2003) [arXiv:hep-th/0301180].

\bibitem{eff}
T.~Battefeld and S.~Watson,
``Effective field theory approach to string gas cosmology,''
arXiv:hep-th/0403075.

\bibitem{kachru}
S.~B.~Giddings, S.~Kachru and J.~Polchinski,
``Hierarchies from fluxes in string compactifications,''
Phys.\ Rev.\ D {\bf 66}, 106006 (2002)
[arXiv:hep-th/0105097].

\bibitem{liam}
S.~Kachru, R.~Kallosh, A.~Linde, J.~Maldacena, L.~McAllister and S.~P.~Trivedi,
``Towards inflation in string theory,''
JCAP {\bf 0310}, 013 (2003)
[arXiv:hep-th/0308055].

\bibitem{stanford}
L.~Kofman, A.~Linde, X.~Liu, A.~Maloney, L.~McAllister and E.~Silverstein,
``Beauty is attractive: Moduli trapping at enhanced symmetry points,''
arXiv:hep-th/0403001.

\bibitem{bagger}
J.~Bagger and I.~Giannakis,
``Higgs mechanism in string theory,''
Phys.\ Rev.\ D {\bf 56}, 2317 (1997)
[arXiv:hep-th/9703202].

\bibitem{mohaupt1}
T.~Mohaupt and M.~Zagermann,
``Gauged supergravity and singular Calabi-Yau manifolds,''
JHEP {\bf 0112}, 026 (2001)
[arXiv:hep-th/0109055].

\bibitem{mohaupt2}
J.~Louis, T.~Mohaupt and M.~Zagermann,
``Effective actions near singularities,''
JHEP {\bf 0302}, 053 (2003)
[arXiv:hep-th/0301125].

\bibitem{supercosmo}
J.~E.~Lidsey, D.~Wands and E.~J.~Copeland,
``Superstring cosmology,''
Phys.\ Rept.\  {\bf 337}, 343 (2000)
[arXiv:hep-th/9909061].

\bibitem{polchinski}
J.~Polchinski,
``String Theory. Vol. 1: An Introduction To The Bosonic String,''
``String Theory. Vol. 2: Superstring Theory And Beyond''.

\bibitem{targetspace}
A.~Giveon, M.~Porrati and E.~Rabinovici,
``Target space duality in string theory,''
Phys.\ Rept.\  {\bf 244}, 77 (1994)
[arXiv:hep-th/9401139].

\bibitem{bv}
R.~H.~Brandenberger and C.~Vafa,
``Superstrings In The Early Universe,''
Nucl.\ Phys.\ B {\bf 316}, 391 (1989).

\bibitem{gubser}
S.~S.~Gubser,
``String production at the level of effective field theory,''
arXiv:hep-th/0305099.

\bibitem{martinec}
A.~E.~Lawrence and E.~J.~Martinec,
``String field theory in curved spacetime and the resolution of spacelike
singularities,''
Class.\ Quant.\ Grav.\  {\bf 13}, 63 (1996)
[arXiv:hep-th/9509149].

\bibitem{me3}
S.~Watson and R.~Brandenberger,
``Linear perturbations in brane gas cosmology,''
JHEP {\bf 0403}, 045 (2004)
[arXiv:hep-th/0312097];\\
S.~Watson,
``UV perturbations in brane gas cosmology,''
arXiv:hep-th/0402015.

\bibitem{preheat}
J.~H.~Traschen and R.~H.~Brandenberger,
``Particle Production During Out-Of-Equilibrium Phase Transitions,''
Phys.\ Rev.\ D {\bf 42}, 2491 (1990);\\
L.~Kofman, A.~D.~Linde and A.~A.~Starobinsky,
``Towards the theory of reheating after inflation,''
Phys.\ Rev.\ D {\bf 56}, 3258 (1997)
[arXiv:hep-ph/9704452].

\bibitem{production}
A.~E.~Lawrence and E.~J.~Martinec,
``String field theory in curved spacetime and the resolution of spacelike
singularities,''
Class.\ Quant.\ Grav.\  {\bf 13}, 63 (1996)
[arXiv:hep-th/9509149];\\
S.~S.~Gubser,
``String production at the level of effective field theory,''
arXiv:hep-th/0305099;\\
J.~J.~Friess, S.~S.~Gubser and I.~Mitra,
``String creation in cosmologies with a varying dilaton,''
arXiv:hep-th/0402156.

\bibitem{peebles}
S.~S.~Gubser and P.~J.~E.~Peebles,
``Structure formation in a string-inspired modification of the cold dark matter
model,''
arXiv:hep-th/0402225.

\end{thebibliography}
\end{document}